\documentclass[prb,twocolumn,superscriptaddress]{revtex4}
\usepackage{graphicx}
\begin{document}

\title[LiBC by Polarized Raman]{
LiBC by Polarized Raman Spectroscopy: Evidence for Lower Crystal
Symmetry?}

\date{\today}

\author{J. Hlinka}
\author{ I. Gregora}
\author{ J. Pokorn\'y}
\affiliation{Institute of Physics ASCR, Praha, Czech Republic}
\author{A. V. Pronin}
\author{A. Loidl}
\affiliation{ Institut f\"{u}r Physik, Universit\"{a}t Augsburg,
 Augsburg, Germany}

\begin{abstract}
The paper presents polarized Raman scattering study on a
few-micron-size crystallite of LiBC with natural faces. The
experiment on as grown sample has revealed a four lattice modes
with frequencies at 1276~cm$^{-1}$, 830~cm$^{-1}$, 546~cm$^{-1}$
and 170~cm$^{-1}$, respectively. The number of observed Raman
lines and their selection rules are incompatible with the
 assumed $\rm D_{6h}$ symmetry. The modes at
1276~cm$^{-1}$ and 170~cm$^{-1}$ correspond to the expected Raman
active modes. In contrast with the superconducting compound
MgB$_2$, the B-C bond stretching mode (at 1276~cm$^{-1}$) has
rather small damping. The two "forbidden" modes (at 830~cm$^{-1}$
and 546~cm$^{-1}$) disappeared after subsequent thermal treatment.
\end{abstract}


\maketitle

Recently there were many attempts to further increase the
remarkably high phase transition temperature toward the
superconducting state in MgB$_2$\cite{nature1} by various chemical
substitutions and doping, but with very little
progress.\cite{Ros02} A new hope was brought by theoretical
proposition that in a related isoelectronic compound, LiBC,
significantly higher phase transition temperatures can be attained
by hole doping produced by Li non-stoichiometry or by field
injection technique\cite{Ros02}. Since the spectroscopic
information on the phonon spectra in these electron-phonon
coupling type superconductors\cite{BCS,Kortus,Bela,An,BOH,Yil} is
obviously important, we have undertaken a basic polarized Raman
scattering characterization of both MgB$_2$\cite{my} and LiBC. The
previous polarized Raman study on MgB$_2$\cite{my} have proven
that the unusually broad spectral feature around 600~cm$^{-1}$
corresponds indeed to the $\rm E_{2g}$ zone center mode. The
present polarized Raman study complements the recent results on
LiBC phonons by infrared spectroscopy\cite{Artem,Bha} and
ab-initio calculations\cite{An02}.

According to Ref.\cite{Wor95}, the LiBC crystals belongs to the
$\rm D_{6h}^4$ ($\rm P6_3/mmc$) space group symmetry, with Li, B
and C atoms in 2a, 2c and 2d Wyckoff positions, respectively. Such
structure is thus similar to MgB$_2$, except for the replacement
of Mg by Li and by replacement of B by C at every second position
along in-plane covalent bonds as well as along the hexagonal axis,
what leads to doubled unit cell along the hexagonal axis. The
factor group analysis at the $\Gamma$-point yields $\rm 2A_{2u} +
2B_{1g}+ B_{2u}$ and $\rm 2E_{1u} + 2E_{2g}+ E_{2u}$ zone center
optic modes with atomic displacements parallel and perpendicular
to the hexagonal axis, respectively. Because of the unit cell
doubling, the internal modes of B-C planes form Davydov-like
doublets: one pair of B-C bond stretching modes ($\rm 1E_{1u} +
1E_{2g}$) and a pair of ring puckering modes ($\rm 1A_{2u} +
1B_{1g}$).
 The remaining six modes can
be understood as "external" modes including a pair of purely Li
modes ($\rm 1E_{2u} + 1B_{1g}$) and four other modes in which the
B-C planes vibrate without deformation. Among all these modes,
only the $\rm E_{2g}$ representation is Raman active, so that only
two first order Raman modes are to be expected in LiBC.

The experiments were carried out at room temperature, using a
Renishaw Raman microscope with 514.5~nm (2.41~eV) argon laser
excitation. The instrument allows measurements of polarized Raman
spectra in back scattering from a spot size down to 1-2 microns in
diameter. To minimize heating of the sample in the laser focus,
the laser power was kept below 1mW. We have used the same, about
10 micron thick platelet sample, as in the recent IR reflectivity
study\cite{Artem}. A
 regular hexagonal flat terrace with a diameter of about
10 microns (see Fig.~1 of the Ref.~\onlinecite{Artem}) in the
corner of the sample confirms\cite{Jung} that the hexagonal axis
is perpendicular to the platelet sample.

The backscattering cross-polarized spectrum measured
 directly from this small  hexagonal face is shown in the lower curve
 of the Fig.~1. As expected, there are two sharp Raman lines.
  Obviously, the higher frequency mode (near
 1176~cm$^{-1}$) corresponds to the $\rm E_{2g}$ bond stretching mode
 (neighbors in B-C planes vibrate in anti-phase, neighbors along
 hexagonal axis vibrate in-phase, Li does not participate). It has
 practically the same frequency as the $\rm E_{1u}$ bond stretching mode
 ( neighbors in B-C planes and neighbors along
 hexagonal axis vibrate in anti-phase),
 seen clearly in the IR reflectivity spectra\cite{Artem}. This shows
 that the Davydov-like splitting is rather small, in agreement
 with the ab-initio calculations\cite{An02} predicting the
 two
 zone center bond stretching modes  at\cite{Bha} 1185~cm$^{-1}$
  and 1194~cm$^{-1}$. Let us note that unlike in
  MgB$_2$,
   both these bond stretching
modes
 in LiBC have rather small damping. Therefore,
 the electron-phonon interaction properties of LiBC are
 probably  closer to AlB$_2$.\cite{Pos02,Ren02}

\begin{figure}
\hspace{0cm} \centerline{\includegraphics[width= 8cm, clip=true]
    {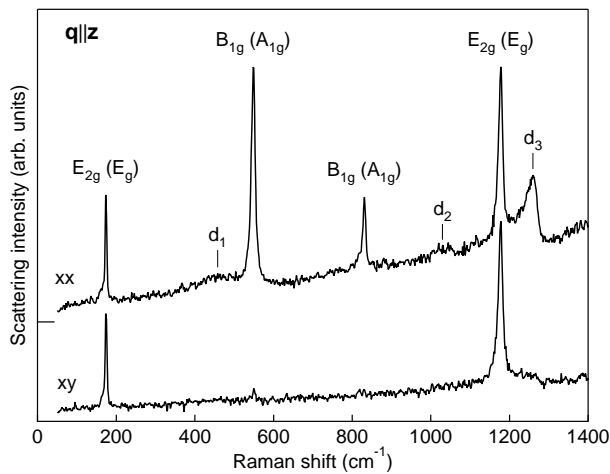}}
 \caption{Polarized Raman spectra taken
from the hexagonal terrace on the LiBC single crystal shown in
 Fig.~1 of Ref.\onlinecite{Artem}. The lower curve corresponds
to the cross-polarized configuration, the upper one, with a
vertical offset shown by the horizontal mark on the vertical axis,
corresponds to the parallel-polarized configuration. Labels are
explained in the text.\label{Fig1}}
\end{figure}

The eigenvector of the second expected $\rm E_{2g}$ mode
represents a shear mode of rigid B-C planes (all atoms in B-C
plane move in the same sense, the next layer vibrates in
anti-phase, Li does not participate). Such mode obviously derives
from the transverse acoustic mode at the A-point of the parent
MgB$_2$ structure, so that its frequency can be estimated from the
velocity of the transverse acoustic modes propagating along the
hexagonal axis. The frequency of the corresponding A-point mode in
MgB$_2$ and AlB$_2$ can be read from Fig. 3 of
Ref.~\onlinecite{Ren02} (200~cm$^{-1}$ and 160~cm$^{-1}$
respectively). Indeed, this estimation nicely agrees with the
frequency of the other mode (near 170~cm$^{-1}$) observed in our
cross-polarized spectra shown in Fig.1.

The upper curve in the Fig.~1 was taken from the exactly same spot
but with parallel polarization.
 In agreement with the expected
symmetry properties of Raman tensors for the $\rm E_{2g}$ modes,
the intensity of the two above discussed modes does not change
significantly. At the same time, several new lines have appeared
in the spectrum. Somewhat higher luminescence background and broad
features d$_1$, d$_2$ and d$_3$ corresponding to the pronounced
bands of LiBC phonon density states (see Fig.~3 of
Ref.~\onlinecite{An02}, Fig.~4 of Ref.~\onlinecite{Bha}) are
rather common for parallel-polarized scattering
configuration\cite{my}. However, the presence of the two
additional sharp lines near 546~cm$^{-1}$ and 830~cm$^{-1}$ is in
a flagrant disagreement with the above symmetry analysis, as if
the actual factor group symmetry were lower than that of the
expected $\rm D_{6h}$ group.

There is no doubt that the 830~cm$^{-1}$ mode is a
"ring-puckering" mode. It is too high frequency for an "external"
mode and the frequency region of bond stretching modes is limited
by the d$_2$ and d$_3$ edges of the phonon density of states.
Moreover, its frequency roughly coincides with ring-puckering
phonon band calculated ab-initio\cite{An02}.

 There are certainly various mechanisms that may be invoked to
 explain the Raman activity of the ring-puckering mode in as-grown LiBC,
 but a  decisive proof for the actual mechanism is in anyway beyond the
 scope of this study. Nevertheless, we would like to point out
 that a rather straightforward explanation can be obtained when
 assuming that the puckering of B-C planes
 is in fact present as a static distortion. In this case the puckering mode, which is
 "frozen in" the parent $\rm P6_3/mmc$ structure, becomes a
 totally symmetric mode of the distorted structure, and such a mode  is
 always Raman active in parallel-polarized geometry.
 Since just four  clear Raman modes have been observed in our spectra,
 we have searched possible candidates for frozen-in modes
 among the zone center
 modes only. As mentioned previously, there are only two such puckering modes
 ($\rm 1A_{2u} + 1B_{1g}$).
We can exclude the $\rm A_{2u}$ mode because it would reduce the
factor
 group symmetry towards $\rm C_{6v}$ (within $\rm C_{6v}$,  the
  $\rm E_{1u}$ should appear together with the two $\rm E_{2g}$ modes
 as 3 $\rm E_{2}$ modes in the cross-polarized spectra, while
  only two modes were observed there). Thus, we are left with a single candidate,
  the $\rm B_{1g}$ mode.
The $\rm B_{1g}$ zone center mode distortion leads to the $\rm
D_{3d}^3$ ($\rm P\bar{3}m 1$) structure, which is compatible with
our experimental data, as we shall show in the following.

  First of all, the  factor group analysis for the hypothetic distorted
   structure yields $\rm 3A_{2u}
+ 2A_{1g}$ and $\rm 3E_{u} + 2E_{g}$ zone center optic modes with
atomic displacements parallel and perpendicular to the trigonal
axis, respectively. They are all either Raman ($\rm 2A_{1g}$ and
$\rm 2E_{g}$) or infrared ($\rm 3A_{2u}$ and $\rm 3E_{u}$) active
modes. To facilitate
  further discussion, from now on we  distinguish  the symmetry adapted
eigenvectors by the symbols of correspondent irreducible
representation in both $\rm D_{6h}$ and $\rm D_{3d}$ factor
groups, the latter being in the parentheses.

The observation shown in the Fig.~1 is in agreement with the Raman
selection rules for the $\rm D_{3d}$ factor group. The pair of
$\rm B_{1g} (A_{g})$ modes should be observable in
parallel-polarized but not in cross-polarized geometry, which
indeed holds for the lines near 546~cm$^{-1}$ and 830~cm$^{-1}$.
The requirement that the pair of $\rm E_{2g} (E_{g})$ modes should
be observable in both cross-polarized and parallel-polarized
geometry with the same intensity holds both for $\rm D_{3d}$ and
$\rm D_{6h}$) symmetry ( and is thus also fulfilled).

Measurements analogous to those shown in the Fig. 1 were
reproduced at several places in the sample, both inside and
outside the small hexagonal terrace, and for different
orientations of incident light polarization axis, all with similar
result. To obtain one more supplementary argument in favor of our
conclusion, we have tried to measure the remaining polarization
components from backscattering on the edges of the sample.
Unfortunately, we have not found well developed flat surfaces
perpendicular to the platelet, so that this measurement was rather
difficult and the phonon propagation direction may not be strictly
perpendicular to the c-axis. Nevertheless, the obtained results
are again in agreement with the proposed assignment within the
$\rm D_{3d}$ factor group. The pair of $\rm E_{2g} (E_{g})$ modes
is present even when one of the polarization axes is parallel to
the hexagonal (trigonal) axis (see "zz" and "zx" spectra shown in
Fig.~2), which is expected within $\rm D_{3d}$ but not allowed in
$\rm D_{6h}$. Similarly, the $\rm B_{1g} (A_{g})$ modes, are
missing in cross-polarized geometry, as required by $\rm D_{3d}$
group Raman selection rules. Finally,
 no frequency shifts which could appear in the case
of polar modes due to the LO-TO splitting were observed, again in
agreement with our above conjecture that the frozen displacement
is not the $\rm A_{2u}$ polar mode.

\begin{figure}
\hspace{0cm} \centerline{\includegraphics[width= 8cm, clip=true]
   {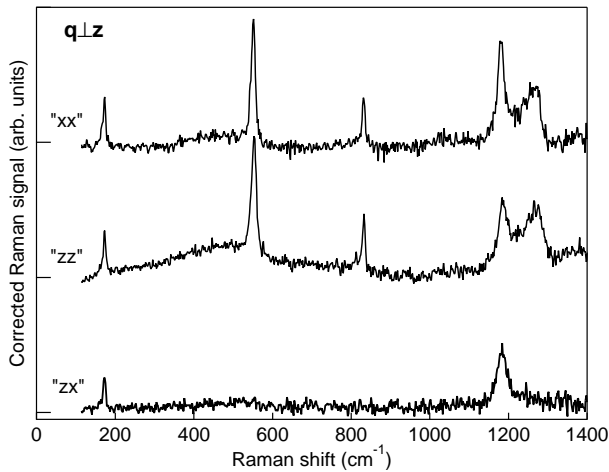}}
 \caption{Polarized Raman spectra taken
from the edge of the LiBC single crystal shown in
 Fig.~1. of Ref.\onlinecite{Artem}. For clarity, luminescence
 background was subtracted, the vertical
offset is shown by the horizontal marks on the vertical axis. The
labels "xx", "zz", "xz" denotes incident and scattered
polarizations (z and x stands for polarization along and
perpendicular to the trigonal axis, respectively.)\label{Fig2}}
\end{figure}

Samples of the same origin were previously studied by infrared
spectroscopy.\cite{Artem} Among the six remaining zone center
optic modes, three have dipole moment along the trigonal axis
($\rm 2A_{2u} (A_{2u}) + 1B_{2u} (A_{2u})$) and three have dipole
moments perpendicular to the trigonal axis ($\rm 2E_{1u} (E_{u}) +
1E_{2u} (E_{u})$). The latter triplet should contribute in the
infrared reflectivity, measured from the large surface of the
investigated sample. This is in agreement with the measurements of
Ref.~\onlinecite{Artem}, which shows a clear B-C stretching mode
near 1180~cm$^{-1}$ (this has to be $\rm E_{1u} (E_{u})$ mode) and
a broad band with two well defined peaks (which could be
tentatively assigned to the remaining
 $\rm E_{1u} (E_{u}) + E_{2u} (E_{u})$ modes).

After completing of the above described investigations, the sample
was placed in a variable-temperature cell (LINKAM) in a hope to
drive the system across a hypothetic phase transformation towards
the prototype P6$_3$/mmc structure by thermal treatment. For this
reason, the sample was first cooled down to about 80 K and then
heated up in steps of about 50~K in order to see the evolution of
the Raman spectrum. There was no clear indication of such a phase
transformation up to 400~K. Unfortunately, in the vicinity of 450
K, a sudden drop-out of electricity stopped the experiment.
Continuation of the experiment was possible two days later. This
time the measurements were done while heating the sample from the
room temperature up to 650~K. To our surprise, the two forbidden
lines were never observed again.

In conclusion, we have observed scattering by the two $\rm E_{2g}$
LiBC Raman active modes ( at 1276~cm$^{-1}$, and 172~cm$^{-1}$,
respectively). The relatively high frequency and small damping of
the stretching mode correlates\cite{An02} with the absence of the
superconductivity in the stoichiometric LiBC. This indicates that
the Raman scattering provides a very useful tool for easy testing
of MgB$_2$ related superconducting materials.

We have found that results of the polarized Raman measurements on
the as-grown LiBC single crystal are incompatible with the assumed
LiBC structure. Most unexpectedly, we have found two other modes
at 830~cm$^{-1}$ and 546~cm$^{-1}$. The former forbidden mode is
probably the $\rm B_{1g}$ B-C puckering mode. The observed results
can be explained assuming lower crystal symmetry, e.g $\rm
P\bar{3}m1$ space group symmetry induced by puckering displacement
of B-C planes. This need not be the correct picture (ab-initio
test calculations have not found such puckered structure
stable\cite{private}), but it nevertheless gives some insight into
the phenomenon.

The two forbidden Raman modes vanished after a subsequent thermal
treatment. The influence of thermal treatment of LiBC based
materials seems to be of essential importance for their
superconductor properties, and the appearance (and disappearance)
of forbidden Raman lines in LiBC samples may be one of the clue in
understanding of the relevant microscopical processes. For
example, B-C order stacking faults and/or Li non-stoichiometry
allows a number of possible scenarios. Obviously, detailed
understanding calls for further investigations. It seems that the
sample is sensitive to heating to temperatures exceeding 400~K. We
plan to perform Raman investigations of the thermal treatment in
LiBC on new samples in future.

\begin{acknowledgements}
This work has been supported by the Grant Agency of the Czech
Republic (Postdoc project 202/99/D066). We have appreciated
 critical reading of the manuscript by J.
Petzelt from Institute of Physics AS CR.
\end{acknowledgements}


\begin{thebibliography} {99}

\bibitem{nature1} J. Nagamatsu, N. Nakagawa, T. Muranaka,
Y. Zenitani and J. Akimitsu, Nature {\bf 410}, 63 (2001).
\bibitem{Ros02} H. Rosner, A. Kitaigorodsky, and W. E. Pickett,
 Phys.Rev.Lett. {\bf 88}, 127001 (2002).

\bibitem{BCS} S. L. Bud'ko, G. Lapertot, C. Petrovic, C. E. Cunningham, N. Anderson,
 and P. C. Canfield, Phys.Rev.Lett. {\bf 86}, 1877 (2001).
\bibitem{Kortus} J. Kortus, I. I. Mazin, K. D. Belashchenko,
V. P. Antropov, and L. L. Boyer, Phys.Rev.Lett. {\bf 86}, 4656
(2001), cond-mat/0101446.
\bibitem{Bela} K. D. Belaschenko, M. van Schilfgaarde and V. P.
Antropov, cond-mat/0102290.
\bibitem{An} J. M. An and W. E. Pickett, Phys.Rev.Lett. {\bf 86}, 4366
(2001), cond-mat/0102391.
\bibitem{BOH} K.-P. Bohnen, R. Heid, and B. Renker,
cond-mat/0103319.
\bibitem{Yil} T. Yildirim et al., to appear in Phys.Rev.Lett.,
cond-mat/0103469.
\bibitem{my} J. Hlinka, I. Gregora, J. Pokorn\'y, A. Plecenik, P. K\'{u}\v{s},
L. Satrapinsky and \v{S}. Be\v{n}a\v{c}ka,
 Phys. Rev. B {\bf 64}, 14503(R)(2001).
\bibitem{Artem} A.V. Pronin, K. Pucher, P. Lunkenheimer, A. Krimmel, and A. Loidl,
 cond-mat/0207299.
\bibitem{Bha}  A. Bharathi, S. Jemima Balaselvi, M. Premila, T.N. Sairam, G.L.N. Reddy,
C.S. Sundar, and Y. Hariharan, cond-mat/0207448.
\bibitem{An02}  J.M. An, S.Y. Savrasov, and W.E. Pickett, cond-mat/0207542.
\bibitem{Wor95}  M. W\"{o}rle, R. Nesper, G. Mair, M. Schwartz, and H.G. von Schnering,
Z. Anorg. Allg. Chem. {\bf 621} 1153,(1995).
\bibitem{Jung} C.U. Jung, J.Y. Kim, P. Chowdhury, Kijoon H.P. Kim,
Sung-Ik Lee, D.S. Koh, N. Tamura, W.A. Caldwell, and J.R. Patel
 cond-mat/0203123.
\bibitem{Pos02} P. Postorino, A. Congeduti, P. Dore, A. Nucara, A.
Bianconi, D. Di Castro, S. De Negri, and A. Saccone,
 Phys.Rev. B {\bf 65}, 020507(R) (2002).
\bibitem{Ren02} B. Renker, K.B. Bohnen, R. Heid, D. Ernst, H. Schober, M. Koza, P. Adelmann,
P.Schweiss, and T. Wolf,
 Phys.Rev.Lett. {\bf 88}, 067001 (2002).
\bibitem{private} W.E. Pickett, private communication


\end{thebibliography}
\end{document}